\documentclass[aps,preprint]{revtex4}%
\usepackage{amsfonts}
\usepackage{amsmath}
\usepackage{amssymb}
\usepackage{graphicx}%
\begin{document}
\title{The modified first laws of thermodynamics of anti-de Sitter and de Sitter space-times}
\author{Deyou Chen}
\email{dchen@cwnu.edu.cn}
\affiliation{Institute of Theoretical Physics, China West Normal University, Nanchong, 637009, China}
\author{Qingyu Gan}
\email{qingyugan@hotmail.com}
\affiliation{Center for Theoretical Physics, College of Physical Science and Technology, Sichuan University, Chengdu, 610064, China}
\author{Jun Tao}
\email{taojun@scu.edu.cn}
\affiliation{Center for Theoretical Physics, College of Physical Science and Technology, Sichuan University, Chengdu, 610064, China}

\begin{abstract}{We modify the first laws of thermodynamics of a Reissner-Nordstrom anti-de Sitter black hole and a pure de Sitter space-time by the surface tensions. The corresponding Smarr relations are obeyed. The cosmological constants are first treated as fixed constants, and then as variables associated to the pressures. For the black hole, the law is written as $\delta E = T \delta S - \sigma\delta A$ when the cosmological constant is fixed, where  $E$ is the Misner-Sharp mass and $\sigma$ is the surface tension. Adopting the varied constant, we modify the law as $\delta E_0 = T \delta S - \sigma_{eff}\delta A +V\delta P$, where $E_0=M-\frac{Q^2}{2r_+}$ is the enthalpy.  The thermodynamical properties are investigated. For the de Sitter space-time, the expressions of the modified laws are different from these of the black hole. The differential way to derive the law is discussed.}
\end{abstract}
\maketitle
\tableofcontents

\section{Introduction}
Thermodynamical properties of anti-de Sitter (AdS) and de Sitter (dS) space-times attract considerable attentions. The research on thermodynamical properties of AdS spaces originated in the seminal work of Hawking and Page. In their work, the existence of the phase transition in the Schwarzschild AdS black hole was first found \cite{HP}. The phase transitions and the thermodynamical stabilities in the various complicated spacetimes were studied in the subsequent work \cite{CG,CG1}.

The cosmological constants were usually seen as the fixed constants. However, they were recently treated as the variables \cite{KRT,KRT1,BD,KM1,KM2,WWX,YS,UTS,UTS1}. The first reason is that the physical constants, such as Newtonian's constant, gauge coupling constants and the cosmological constant arise as vacuum expectation values and vary in more fundamental theories \cite{KM1}. The second reason is reconciling the inconsistencies between the first law of black hole thermodynamics and the Smarr relation derived by the scaling method. Associating the cosmological constant with the pressure $P=-\frac{\Lambda}{8\pi}$, Kastor et al first derived the modified first law of thermodynamics and the corresponding Smarr relation of the static AdS space-time by the geometric approach \cite{KRT}. The key ingredient is the two-form potential for the Killing field. The new term $VdP$ appeared in the law and the black hole mass was naturally interpreted as the enthalpy. In \cite{BD}, Dolan obtained the equation of state for the AdS black hole and discussed the analogy with that of the Van der Waals system. This work got further development and some interesting results were found in \cite{KM1}. Kubiznak and Mann revealed the identification of the Reissner-Nordstrom anti-de Sitter (RN-AdS) black hole with the liquid-gas system in the extended phase space by investigating the behavior of the Van der Waals system and that of the Gibbs free energy in the fixed charge ensemble. The modified first law in the extended phase space was gotten as

\begin{equation}
\delta M = T\delta S +\Phi \delta Q+ V\delta P. \label{eq1.1}
\end{equation}

\noindent where $\Phi$ is the electromagnetic potential at the horizon and $V$ is the black hole volume. The extensions to other complicated space-times are referred to \cite{HV,CCL,ZZW,BPD,WL,BMS,GKY,AOS,BD5,BD2,BD3}. In the AdS/CFT correspondence, the variation of the cosmological constant is associated to that of the degrees of freedom in the gauge theory. Treating the constant as the number of colors in the field, Dolan calculated the chemical potential conjugate to the number and studied the thermodynamics of the $AdS_5\times S^5$ space-times in \cite{BD1}. He found that the potential in the high temperature is negative and decreases as the increase of the temperature. When the temperature is lowered below the Hawking-Page temperature, the potential approaches to zero. This work was further discussed in \cite{ZCY,ZCY1}. It was found that the chemical potential becomes positive more easily due to the existence of the charge in RN-AdS black holes. The positive cosmology constant was first seen as a variable in \cite{WWX}. Taking into account the variable cosmology constant, Wu and Sekiwa studied the thermodynamics of the dS spacetimes, respectively \cite{WWX,YS}. Subsequently, the constant was also treated as the pressure and the thermodynamic properties of the charged and rotating de Sitter black holes in the extended phase spaces were studied in \cite{DKK,ZZM,ZZM1,KS}.

On the other hand, Einstein's equations can be written as a thermodynamical identity \cite{TP,RGC,HKM}. It was first proved in the spherically symmetric space-time and the equations were written in the form of \cite{TP}

\begin{equation}
\delta E = T\delta S- P\delta V, \label{eq1.2}
\end{equation}

\noindent where $\delta V$ is the change of space volume and $P$ is the pressure provided by the source of Einstein's equations. This result could also be applicable for stationary spacetimes and other gravity theories \cite{TP1,TP2,TP3}. Recently, this work was extended to the Kerr black hole \cite{HKM}. The horizon energy $E$ of the black hole was identified as its Misner-Sharp mass, and then the angular momentum was $Ea$. The horizon radius $r_+$ and the rotation parameter $a$ were independent variables. These identifications are different from the traditional concept where $a$ is a fixed constant. From the relation between the entropy and the horizon area, the surface tension $\sigma$ and the work term $\sigma \delta A$ were gotten. Thus, Einstein's equations were rewritten as the modified first law of black hole thermodynamics

\begin{equation}
\delta E = T\delta S  + \Omega \delta J -\sigma \delta A.  \label{eq1.3}
\end{equation}

In this paper, we modify the first laws of thermodynamics of the RN-AdS black hole and the pure dS space-time by the surface tensions. We first treat the cosmological constant as a fixed constant and derive the smodified first law of the RN-AdS black hole. The corresponding Smarr relation is obeyed. Then we revisit the law by associating the cosmological constant with the pressure. The effective surface tension and the modified law are gotten. The corresponding Smarr relation is also satisfied. Thermodynamics of dS space-times is more complex than that of AdS space-times. The reason is that the non-equilibrium state exists in this region between the black hole horizon and the cosmological horizon due to the different temperatures at the different horizons. Meanwhile, the asymptotic mass is difficult to be defined since the Killing vector is timelike outside the black hole horizon. To overcome these problems, some effective ways were put forward \cite{UTS1,AC,YS1,CJS,WH}. Here we modify the first law at the cosmological horizon of the pure dS space-time where only a cosmological horizon exists. In the dS space-time, the energy is measured as the negative value. We identify the negative Misner-Sharp mass as the horizon energy. The modified first laws and the corresponding Smarr relations are gotten when the cosmological constant is fixed and  varied, respectively. The modified first laws at the black hole horizon and the cosmological horizon take on different forms.

The rest is outlined as follows. In the next section, using the expressions of the temperature and the radial Einstein's equations, we derive the surface tension and the modified first laws of thermodynamics of the RN-AdS black hole, where the cosmological constant is seen as a fixed constant and a variable, respectively. In section 3, the surface tension and the modified first laws at the cosmological horizon of the dS space-time are investigated. Section 4 is devoted to our discussion and conclusion.

\section{Modifying the first law of thermodynamics at the black hole horizon}

\subsection{The first law with the fixed cosmology constant}

The metric of the RN-AdS black hole is given by

\begin{equation}
ds^2 = -\Delta dt^2 + \frac{1}{\Delta} dr^2 +r^2(d\theta^2 +sin^2{\theta}d{\phi}^2), \label{eq2.1}
\end{equation}

\noindent with the electromagnetic potential $A_{\mu} =\frac{Q}{r}dt$, where

\begin{equation}
\Delta = 1- \frac{2M}{r} + \frac{Q^2}{r^2} + \frac{r^2}{l^2}, \label{eq2.2}
\end{equation}

\noindent $2M = r_+ + \frac{r_+^3}{l^2}+\frac{Q^2}{r_+}$. $l$ is related to the cosmological constant $\Lambda$ as $l^2= - \frac{3}{\Lambda}$. The black hole horizon is located at the largest root $r_+$ obtained from $\Delta =0$. $M$ and $Q$ are the physical mass and charge, respectively. There are several ways to define the mass. Here we adopt the definition of the Misner-Sharp mass to investigate the thermodynamical properties of the black hole \cite{MS1}. The explicit form of the mass was presented in \cite{MS2,SAH,HZ}. For the RN-AdS black hole, its mass is gotten as $E=M-\frac{Q^2}{2r_+}-\frac{r_+^3}{2l^2}$. Introducing the expression of the physical mass $M$ yields a concise expression

\begin{equation}
E=\frac{r_+}{2}. \label{eq2.3}
\end{equation}

\noindent The black hole entropy and temperature are

\begin{eqnarray}
S & = & \frac{A}{4}=\pi r_+^2, \nonumber\\
T & = & \frac{\Delta^{\prime}(r_+)}{4\pi} = \frac{1}{4\pi r_+}\left(1+\frac{3r_+^2}{l^2}-\frac{Q^2}{r_+^2}\right),\label{eq2.4}
\end{eqnarray}

\noindent respectively, where $\Delta^{\prime}(r_+)=\frac{\partial\Delta}{\partial r}\mid _{r=r_+}$. The value of $T$ is determined by $r_+$ and $Q$. The Figure 1 describes the variation relation between $T$ and $r_+$ at the fixed values $Q$ and $l=1$. When the horizon radius is small, the temperatures are different for the different $Q$. Finally, the temperatures approach to the same value at the large radius.

\begin{figure}[tb]
\begin{centering}
\includegraphics[scale=.7]{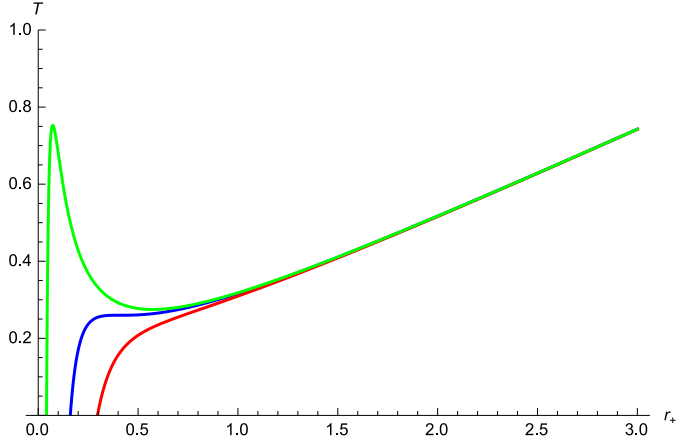}
\par\end{centering}
\caption{The red, blue and green curves correspond to $Q = 1/3$, $Q = 1/6$ and $Q = 1/24$, respectively.}%
\label{fig:1}%
\end{figure}

Einstein's equations can be written as a thermodynamic identity. To investigate the thermodynamics and the surface tension at the black hole horizon, we first calculate the radial Einstein equation at the horizon. The equation is

\begin{equation}
G_r^r\mid_{r_+} = 8\pi T_r^r\mid_{r_+}= \frac{r_+\Delta^{\prime}(r_+)-1}{r_+^2}. \label{eq2.5}
\end{equation}

\noindent Compared Eq. (\ref{eq2.4}) and Eq. (\ref{eq2.5}), the black hole temperature is rewritten as

\begin{equation}
T = \frac{8\pi r_+^2 T_r^r\mid_{r_+} +1}{4\pi r_+}. \label{eq2.6}
\end{equation}

\noindent Carrying out differential on the entropy yields $\delta S = 2\pi r_+ \delta r_+$, where $r_+$ is seen as an independent variable. Multiplying by $\delta S$ on the both sides of  Eq. ({\ref{eq2.6}}), we get

\begin{equation}
T \delta S= 2 r_+ T_r^r\mid_{r_+}\delta S +\frac{\delta r_+}{2}. \label{eq2.7}
\end{equation}

\noindent The first term on the right hand side (rhs) in the above equation is relied on the matter, while the second term can be identified as the differential form of the Misner-Sharp mass, namely, $\delta E= \frac{1}{2}\delta r_+$. Thus, Eq. ({\ref{eq2.7}}) is written as

\begin{equation}
\delta E = T \delta S - \sigma\delta A, \label{eq2.8}
\end{equation}

\noindent where $\sigma = \frac{ r_+ T_r^r\mid_{r_+}}{2}= \frac{1}{16\pi}(-\frac{Q^2}{r_+^3}+\frac{3r_+}{l^2})$ describes the surface tension at the horizon and $A=4S$ is the horizon area. Eq. ({\ref{eq2.8}}) is the modified first law of thermodynamics at the black hole horizon. The value of the surface tension is dependent on both of the charge and the cosmological constant. When $3r_+^4=Q^2l^2$, the surface tension is zero. The positive surface tension is gotten for $3r_+^4>Q^2l^2$, while it is negative when $3r_+^4<Q^2l^2$. The corresponding Smarr relation

\begin{equation}
E = 2T S - 2\sigma A, \label{eq2.9}
\end{equation}

\noindent is satisfied by using the expressions of $E$, $T$, $S$, $A$ and $\sigma$. Clearly, the first law in Eq. ({\ref{eq2.8}}) is different from that gotten in \cite{KM1,KM2}. The reason is that the Misner-Sharp mass was introduced as the energy and the temperature was expressed by the radial Einstein equation. When the ADM mass expresses the energy, Eq. (\ref{eq2.8}) is reduced to Eq. (3.9) in \cite{KM1}.

The Gibbs free energy is

\begin{equation}
G = E- T S+ \sigma A. \label{eq2.10}
\end{equation}

\noindent Using the expressions of $E,T,S,\sigma$ and $A$, we get $G=\frac{r_+}{4}$. The above equation obeys the differential form

\begin{equation}
\delta G = -S\delta T + A\delta \sigma. \label{eq2.11}
\end{equation}

\noindent If we order the effective temperature be $T_{eff}=T-4\sigma=\frac{1}{4\pi r_+}$, Eq. ({\ref{eq2.8}}) is then written as $\delta E = T_{eff} \delta S $. The corresponding Gibbs free energy is $G_{eff} = E- T_{eff} S $ which obeys $\delta G_{eff} = -S\delta T_{eff}$. The $G-T$ and  $G_{eff}-T_{eff}$ diagrams are displayed in the figures 2 and 3, respectively. In the figure 2, we find that the the Gibbs free energies are different at the different surface tensions when the temperature is very low. Finally, the energies approach to the same value with the increase of the temperature.

\begin{figure}[tb]
\begin{centering}
\includegraphics[scale=.7]{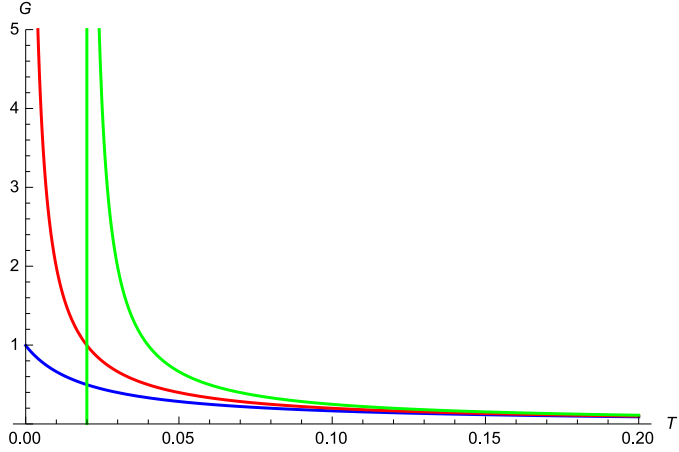}
\par\end{centering}
\caption{The blue, red and green curves correspond to $\sigma = -0.005$, $\sigma = 0$ and $\sigma = 0.005$, respectively.}%
\label{fig:2}%
\end{figure}

\begin{figure}[tb]
\begin{centering}
\includegraphics[scale=.7]{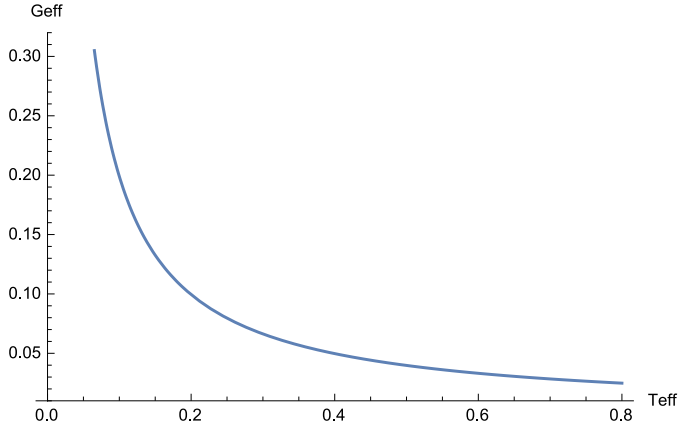}
\par\end{centering}
\caption{The diagram shows that the effective Gibbs free energy decreases with the increase of the effective temperature.}%
\label{fig:3}%
\end{figure}

\subsection{The pressure}

In this subsection, we discuss the pressure at the horizon. Using the relation between the area and the volume, $V=\frac{Ar_+}{3}$, yields

\begin{equation}
\sigma\delta A = P\delta V, \label{eq3.1}
\end{equation}

\noindent where $P= T_r^r\mid_{r_+}= \frac{1}{8\pi}(-\frac{Q^2}{r_+^4}+\frac{3}{l^2})$ is the pressure. Thus, the modified law becomes $\delta E = T \delta S - P\delta V $. The positive (zero and negative) pressures correspond to the positive (zero and negative) surface tensions. Using the expressions of the pressure and the temperature, we get

\begin{equation}
P= \frac{T}{2r_+} - \frac{1}{8\pi r_+^2}. \label{eq3.3}
\end{equation}

\noindent In the Van der Waals equation $P= \frac{kT}{v - b} - \frac{a}{v^2}$, $v$, $T$ and $k$ are the specific volume, the temperature and the Boltzmann constant, respectively. $a$ is a positive constant described the attraction of molecules in the system, while $b$ is the size of molecules. From Eq. (\ref{eq3.3}), we get $a=\frac{1}{2\pi}$ and $b=0$, which show that the molecules are ideal point particles and their interaction force is the attractive force. $v=2r_+$ is identified as the specific volume. Then Eq. (\ref{eq3.3}) is written as

\begin{equation}
P= \frac{T}{v} - \frac{1}{2\pi v^2}. \label{eq3.4}
\end{equation}

\noindent When $Q=0$, the metric (\ref{eq2.1}) is reduced to the Schwarzschild AdS black hole and the pressure gotten as $P=-\frac{\Lambda}{8\pi }$ is positive always. When $\Lambda=0$, it describes the RN black hole and the negative pressure $P=-\frac{Q^2}{8\pi r_+^4 }$ is naturally obtained. The pressure is zero when $Q=\Lambda=0$, which is full inconsistence with that of the Schwarzschild black hole described the vacuum solution with no pressure.

The Gibbs free energy is $G=E-TS+PV$, which obeys $\delta G= -S \delta T +V\delta P$. The figure 4 depicts the Gibbs free energy varying with the temperature at the fixed pressures. It shows that a characteristic cusp exists when $P>0$.

\begin{figure}[tbp]
\begin{centering}
\includegraphics[scale=0.65]{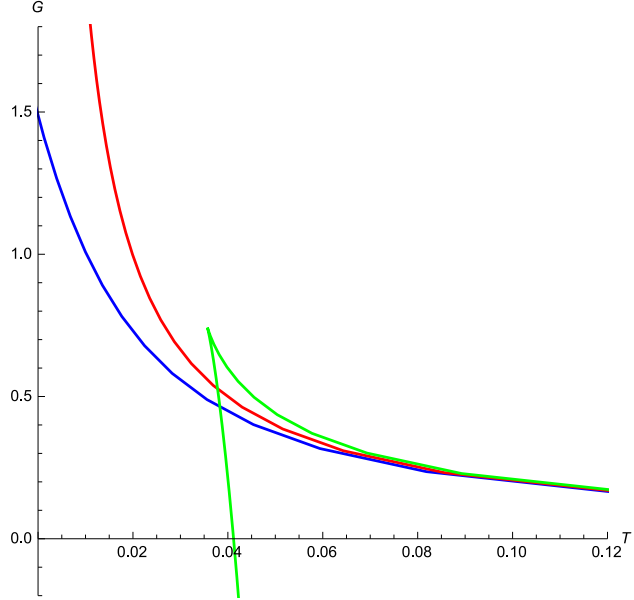}
\par\end{centering}
\caption{The curves from left to right correspond to $P=-0.002$, $P=0$ and $P=0.002$, respectively.}%
\label{fig:4}%
\end{figure}

\begin{figure}[tbp]
\begin{centering}
\includegraphics[scale=0.65]{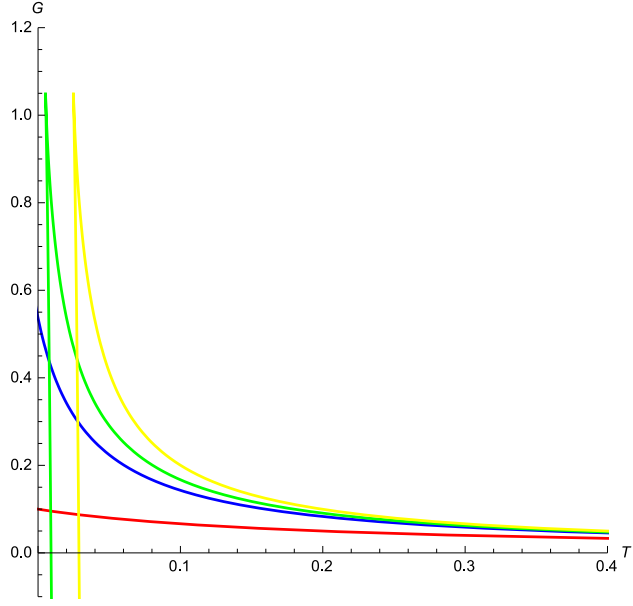}
\par\end{centering}
\caption{The blue, red, green and yellow curves correspond to $\sigma = -0.01$, $\sigma = -0.05$, $\sigma = -0.005$  and $\sigma = -0.0001$, respectively.}%
\label{fig:5}%
\end{figure}

\begin{figure}[tbp]
\begin{centering}
\includegraphics[scale=0.65]{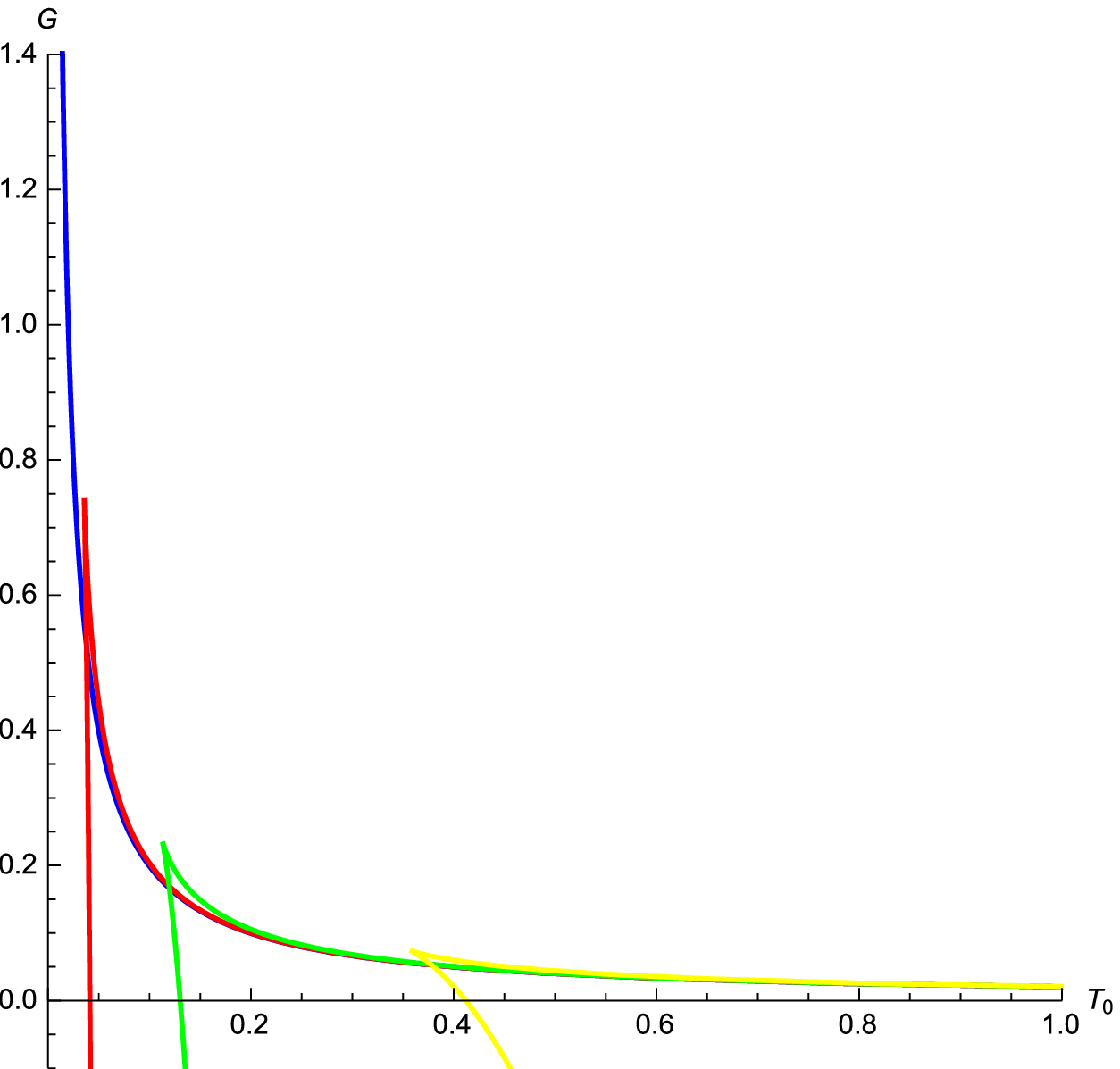}
\par\end{centering}
\caption{The curves from left to right correspond to $P=0$, $P=0.002$, $P=0.02$ and $P=0.2$, respectively.}%
\label{fig:6}%
\end{figure}

\subsection{The first law with the varied cosmology constant}

In the recent work, the cosmological constant was seen as an varied constant associated to the pressure. In this subsection, we adopt the varied cosmology constant and make further efforts to explore the thermodynamics of the RN-AdS black hole. Rewrite the first term on rhs of Eq. ({\ref{eq2.7}}) as

\begin{equation}
2 r_+ T_r^r\mid_{r_+}\delta S = -\frac{Q^2}{4\pi r_+^3}\delta S +\frac{3r_+}{4\pi l^2}\delta S =  -\frac{Q^2}{16\pi r_+^3}\delta A +\frac{3}{8\pi l^2}\delta V. \label{eq3.5}
\end{equation}

\noindent The last equal sign in the above equation was gotten by using the the relation among the entropy, horizon area and volume. The first term relies on the charge and the second term depends on the cosmological constant. We introduce the definition of the pressure $P = -\frac{\Lambda}{8\pi}$ and let $\sigma_{eff} = -\frac{Q^2}{16\pi r_+^3}$. Eq. ({\ref{eq2.7}}) becomes

\begin{eqnarray}
\delta E = T \delta S - \sigma_{eff}\delta A -P\delta V, \label{eq3.6}
\end{eqnarray}

\noindent which can be rewritten as

\begin{eqnarray}
\delta E_0 = T \delta S - \sigma_{eff}\delta A +V\delta P, \label{eq3.7}
\end{eqnarray}

\noindent by using the transformation $P\delta V = \delta(PV)-V\delta P$, where $E_0= E+PV=M-\frac{Q^2}{2r_+}$ is identified as the enthalpy. Further analysis shows $E_0$ is just the Misner-Sharp mass of the RN black hole. Taken into account the expressions of $T,S,\sigma, A,P$ and $V$, the corresponding Smarr relation

\begin{eqnarray}
E_0=2(TS-\sigma_{eff} A  -PV),\label{eq3.8}
\end{eqnarray}

\noindent is also satisfied. Therefore, Eq. ({\ref{eq3.7}}) can also been seen as the modified first law of thermodynamics of the RN-AdS black hole. It is an important result in this paper. Due to the existence of the cosmological constant as the variable, there is an additional worked term $V\delta P$ in the law and the total energy is $M-\frac{Q^2}{2r_+}$. Thus, $\sigma_{eff}$ is seen as the effective surface tension. Using the expressions of $T$ and $P$, we get

\begin{equation}
P= \frac{T}{2r_+} - \frac{1}{8\pi r_+^2} + \frac{Q^2}{8\pi r_+^4}. \label{eq3.9}
\end{equation}

\noindent Compared with the Van der Waals equation, the properties of the RN-AdS black hole was discussed in detail in \cite{KM1}. The difference is that the total energy is defined as $M-\frac{Q^2}{2r_+}$ in this subsection, while that is the physical mass $M$ in \cite{KM1}. The corresponding Gibbs free energy is $G=E-TS+\sigma_{eff} A+PV$. The $G-T$ diagram at the fixed $\sigma$ and $P=0.001$ is depicted in the figure 5. It shows that the characteristic cusps corresponded to the same Gibbs free energy when the surface tension is less than a certain value.

Eq. ({\ref{eq3.7}}) can be furthermore written as

\begin{eqnarray}
\delta E_0 = T_0 \delta S  +V\delta P, \label{eq3.10}
\end{eqnarray}

\noindent which obeys the corresponding Smarr relation

\begin{eqnarray}
E_0=2(T_0S- PV),  \label{eq3.11}
\end{eqnarray}

\noindent  where $T_0=T- 4\sigma_{eff}=  \frac{1}{4\pi r_+}\left(1+\frac{3r_+^2}{l^2}\right)$. Therefore, Eq. ({\ref{eq3.10}}) can also be regarded as the modified first law of the black hole thermodynamics. Now the Gibbs free energy is gotten as $G=E-T_0S+PV$. The $G-T_0$ diagram is described in the figure 6. We find that the position of the characteristic cusp depends on the value of the pressure.

\section{Modifying the first law of thermodynamics at the cosmological horizon}

The research on thermodynamics of dS space-times is more complex than that of AdS space-times. In a dS space-time where a black hole horizon and a cosmological horizon exist at the same time, the temperatures at the different horizons are different. Therefore, there is a non-equilibrium state in this space-time. To discuss thermodynamical properties in the equilibrium state, one can treat the black hole horizon and the cosmological horizon as two independent thermodynamical systems \cite{UTS1,AC,YS1,CJS,WH}. Besides, one can adopt a global view to construct the globally effective temperature and other effective thermodynamic quantities.

In this section, we investigate the thermodynamics of the pure dS space-time by the surface tension. The modified first laws of thermodynamics at the cosmological horizon are gotten. The pure dS space-time only contains a cosmological horizon. The metric is given by

\begin{equation}
ds^2 = -F(r) dt^2 + \frac{1}{F(r)} dr^2 +r^2(d\theta^2 +sin^2{\theta}d{\phi}^2), \label{eq4.1}
\end{equation}

\noindent where $F(r)=1- \frac{r^2}{L^2}$, $L^2=\frac{3}{\Lambda}$ and $\Lambda$ is the cosmological constant. The cosmological horizon is located at $r_{C} = L$ derived from $F(r)=0$. The entropy is $S=\frac{A}{4}=\pi r_C^2$. The temperature at the horizon is

\begin{equation}
T=\frac{\mid F^{\prime}(r_C) \mid}{4\pi}= \frac{1}{2\pi L}. \label{eq4.2}
\end{equation}

\noindent In the above equation, $F^{\prime}(r_C)=\frac{\partial F(r)}{\partial r}\mid _{r=r_C} = -\frac{2}{L}$. Since the temperature should be positive, we used the absolute value. Adopting the method in \cite{SAH}, we get the Misner-Sharp mass as

\begin{equation}
M= \frac{r_C}{2}. \label{eq4.3}
\end{equation}

The horizon thermodynamics are related to Einstein's equations. To discuss the surface tension and the thermodynamics, we first calculate the radial Einstein's equation and get

\begin{equation}
G_r^r\mid_{r_C} = 8\pi T_r^r\mid_{r_C}= \frac{r_C F^{\prime}(r_C)-1}{r_C^2}. \label{eq4.4}
\end{equation}

\noindent Solving the above equation yields

\begin{equation}
F^{\prime}(r_C) = 8\pi r_C T_r^r\mid_{r_C}+\frac{1}{r_C}. \label{eq4.5}
\end{equation}

\noindent Moving the first term on rhs in the above equation to the left hand side (lhs) and then multiplying by $-\frac{\delta S}{4\pi}$ on the both sides, we get
\begin{equation}
-\frac{F^{\prime}(r_C)}{4\pi}\delta S + 2 r_C T_r^r\mid_{r_C}\delta S = -\frac{\delta r_C}{2}. \label{eq4.6}
\end{equation}

\noindent Here we have used the differential expression $\delta S = 2\pi r_C \delta r_C$ to derive the term on rhs of Eq. (\ref{eq4.6}). It is clearly that $-\frac{F^{\prime}(r_C)}{4\pi}$ is the value of the temperature at the cosmological horizon. So the first term on lhs is in of the form $T dS$. $ \delta M =\frac{1}{2}\delta r_C$ expresses the change of the Misner-Sharp mass in the universe. In dS space-times, the energy of an object is measured as a negative value, rather than its mass \cite{TP,SDD,DK3,VBBM}. We identify the energy $E$ as the negative Misner-Sharp mass, and then get $\delta E = - \delta M$. Thus, the term on rhs of the above equation denotes the change of the energy. The term $2 r_C T_r^r\mid_{r_C}\delta S$ can be written in of the form $-\sigma \delta A$ or $\sigma \delta A$ by using the relation between the entropy and the horizon area. The expression $-\sigma \delta A$ yields the same form as in Eq. (\ref{eq2.8}). Considering the property of the dS space-time, we adopt the expression $\sigma \delta A$ and get $\sigma = \frac{1}{2} r_C T_r^r\mid_{r_C}=-\frac{3}{16\pi r_C}$. Therefore, Eq. (\ref{eq4.6}) is rewritten as

\begin{equation}
\delta E= T \delta S  +\sigma \delta A, \label{eq4.7}
\end{equation}

\noindent which is the modified first law of thermodynamics at the cosmological horizon and $\sigma$ is the surface tension. The inner space of a black hole horizon is an inaccessible region surrounded by the horizon. On the contrary, the outer space of the cosmological horizon is an inaccessible region where the physics isn't known. To study the property of the cosmological horizon, we only recur to the discussion of the inner space-time (normal region) of the cosmology. When a particle is absorbed by the black hole, the entropy, the horizon area and the volume of inaccessible region increase. In the dS space-time, when a particle entries the inaccessible region, the entropy, horizon area and volume of the cosmology decrease. The corresponding quantities $\delta S$, $\delta A$ and $\delta V$ are negative. Therefore, we should beware of the sign in the calculation. The Smarr relation,

\begin{equation}
E= 2(T S  +\sigma A), \label{eq4.8}
\end{equation}

\noindent corresponding to Eq. (\ref{eq4.7}) is also obeyed. On the other side, the modified first law gives an interpretation of the negative energy in the dS space-time.

The Gibbs free energy is

\begin{equation}
G = E- T S - \sigma A. \label{eq4.9}
\end{equation}

\noindent It is found that the Gibbs free energy is $G=-\frac{1}{4}r_C$ and there is no critical phenomenon appeared. Its differential expression is $\delta G = -S\delta T - A\delta \sigma$.

To make further efforts to discuss the thermodynamics of the dS space-time, we rewrite the second term on lhs of Eq. (\ref{eq4.6}) as

\begin{equation}
 T_r^r\mid_{r_C}\delta \left(\frac{4\pi r_C^3}{3}\right)= P \delta V, \label{eq4.10}
\end{equation}

\noindent where $P=T_r^r\mid_{r_C}=-\frac{\Lambda}{8\pi}$ denotes the pressure. When the cosmology radiates a particle, the volume of the inaccessible region increases, while the volume (normal region) and horizon area of the cosmology decreases. The quantity of increase equals to that of decrease. Thus, Eq. (\ref{eq4.6}) is written as

\begin{equation}
\delta E= T \delta S  + P\delta V. \label{eq4.11}
\end{equation}

\noindent Treating the cosmological constant as a variable associated to the pressure $P=-\frac{\Lambda}{8\pi}$ and ordering $E_0=E-PV$ yield the modified law

\begin{equation}
\delta E_0= T \delta S  - V\delta P, \label{eq4.12}
\end{equation}

\noindent where $E_0$ denotes the total energy of the space-time. This result is full in consistence with that derived by other method \cite{DKK}. We find $E_0 = 0$ from the expressions of $E$, $P$ and $V$. This agree with the zero mass of the pure dS space-time. The corresponding Smarr relation

\begin{equation}
E_0= 2( TS + P V), \label{eq4.13}
\end{equation}

\noindent is obeyed.

In this section, the modified first laws of thermodynamics (\ref{eq4.7}) and (\ref{eq4.12}) were gotten. Clearly, the expressions of the laws at the cosmological horizon of the dS space-time are different from these at the black horizon of the RN-AdS black hole.

\section{Discussion and Conclusion}

In fact, the first law can be directly derived by the differential. Treating the cosmological constant as a variable, we carry out differential on Eq. ({\ref{eq2.2}}) at the black hole horizon and get

\begin{equation}
dM = \left(\frac{M}{r_+}-\frac{Q^2}{r_+^2}-\frac{\Lambda r_+^2}{3}\right)dr_+ +\frac{Q}{r_+}dQ-\frac{r_+^3}{6}d\Lambda, \label{eq5.1}
\end{equation}

\noindent where $r_+$, $M$  and $Q$ are seen as variables and $\Phi=\frac{Q}{r_+}$ is the electromagnetic potential at the horizon. Let the cosmological constant be associated with the pressure $P = -\frac{\Lambda}{8\pi}$ and its conjugate quantity be $V=\frac{4\pi r_+^3}{3}$. It is found that Eq. ({\ref{eq5.1}) is just the expression of the modified first law in Eq. ({\ref{eq1.1}). Once again, carrying out differential on $F=1- \frac{r^2}{L^2}$ at the cosmological horizon yields

\begin{equation}
 0= dr_C + \frac{r_C^3}{6} d\Lambda. \label{eq5.2}
\end{equation}

\noindent Let the positive cosmology constant be also associated to the pressure $P = -\frac{\Lambda}{8\pi}$. Thus Eq. ({\ref{eq5.2}}) takes on the form of

\begin{equation}
0 = T\delta S -V\delta P. \label{eq5.4}
\end{equation}

\noindent This result is full in accordance with that of Eq. ({\ref{eq4.12}), where the total energy $E_0$ is zero.

In this paper, using the radial Einstein's equations at the black hole horizon and the cosmological horizon, we derived the modified first laws of thermodynamics of the RN-AdS and pure dS space-times by the surface tensions. The corresponding Smarr relations were satisfied. The cosmological constants were first fixed, and then were seen as the variables associated to the pressures. For the RN-AdS black hole, the surface tension can be positive, zero and negative, and its value is determined by $r_+$, $Q$ and $l$. Using the relation between the entropy and the horizon area, we got the effective temperature. Considering that the horizon area and the volume are associated with each other, we replaced the area with the volume and got the effective pressure. However, there is somewhat flaw when we eliminated the surface tension in terms of the pressure by introducing the volume. Namely, based on the above replacement, the differential form of the first law could be gotten, while the corresponding Smarr relation isn't obeyed. This phenomenon was also found and explained in \cite{HKM}. In fact, the pressures are usually inconsistent after the surface effects of the curved surface are taken into account. To make further efforts to investigate the thermodynamics, we treated the cosmological constant as the pressure $P =-\frac{\Lambda}{8\pi}$ and rewrote the modified first law as Eq. ({\ref{eq3.7}}). The effective surface tension was gotten. The relation between the Gibbs free energy and the temperature was discussed. For the dS space-time, we identified the energy as the negative Misner-Sharp mass and obtained the modified first laws of thermodynamics (\ref{eq4.7}) and (\ref{eq4.12}). In Eq. (\ref{eq4.6}), the term $2 r_C T_r^r\mid_{r_C}\delta S$ was expressed as $\sigma \delta A$. If we express it as $-\sigma \delta A$, the derived first law takes on the same form as Eq. ({\ref{eq2.8}}) and obeys the corresponding Smarr relation.

\begin{acknowledgments}
We would like to thank Professors S.Q. Wu, H. Yang and P. Wang for their useful discussions. This work is supported by the National Natural Science Foundation of China (Grant No. 11205125), by Sichuan Province Science Foundation for Youths (Grant No. 2014JQ0040) and by the Innovative Research Team in China West Normal University (Grant No. 438061).
\end{acknowledgments}

\end{document}